\documentclass[pra,aps,twocolumn,10pt,showpacs,groupedaddress,superscriptaddress,floatfix]{revtex4-1}
\usepackage{amsmath,amsfonts,amssymb,graphics,graphicx,epsfig,color,times}
\usepackage[utf8x]{inputenc}
\usepackage{bbm, dsfont} 
\usepackage{subfigure}
\usepackage{hyperref}
\usepackage{mathrsfs}
\usepackage{verbatim}
\begin{document}
\title{Spin-incoherent Luttinger liquid of one-dimensional spin-1 Tonks-Girardeau Bose gas: Spin-dependent properties}
\author{H. H. Jen}
\affiliation{Institute of Physics, Academia Sinica, Taipei 11529, Taiwan}
\author{S.-K. Yip}
\affiliation{Institute of Physics, Academia Sinica, Taipei 11529, Taiwan}
\affiliation{Institute of Atomic and Molecular Sciences, Academia Sinica, Taipei 10617, Taiwan}

\date{\today}

\renewcommand{\r}{\mathbf{r}}
\newcommand{\f}{\mathbf{f}}

\def\be{\begin{align}}
\def\ee{\end{align}}
\def\bea{\begin{eqnarray}}
\def\eea{\end{eqnarray}}
\def\ba{\begin{array}}
\def\ea{\end{array}}
\def\bdm{\begin{displaymath}}
\def\edm{\end{displaymath}}
\def\red{\color{red}}

\begin{abstract}
Spin-incoherent Luttinger liquid (SILL) is a different universal class from the Luttinger liquid.\ This difference results from the spin incoherence of the system when the thermal energy of the system is higher than the spin excitation energy.\ We consider one-dimensional spin-$1$ Bose gas in the SILL regime and investigate its spin-dependent many-body properties.\ In Tonks-Girardeau limit, we are able to write down the general wave functions in a harmonic trap.\ We numerically calculate the spin-dependent (spin-plus, minus, and $0$) momentum distributions in the sector of zero magnetization which allows to demonstrate the most significant spin-incoherent feature compared to the spinless or spin-polarized case.\ In contrast to the spinless Bose gas, the momentum distributions are broadened and in the large momentum limit follow the same asymptotic $1/p^4$ dependence but with reduced coefficients.\ While the density matrices and momentum distributions differ between different spin components for small $N$, at large $N$ they approach each other.\ We show these by analytic arguments and numerical calculations up to $N$ $=$ $16$.
\end{abstract}
\maketitle
\section{Introduction}

A plethora of studies on one-dimensional (1D) quantum systems \cite{Giamarchi2004, Haldane1981} of gaseous atoms thrive recently due to the experimental achievements of 1D confined bosons \cite{Paredes2004, Kinoshita2004, Haller2009}.\ Many studies focus on the ground state properties of spinless bosons \cite{Girardeau2001, Papenbrock2003} such as spatial and momentum distributions \cite{Minguzzi2002, Olshanii2003, Xu2015}, quantum magnetism in a spinful Bose gas \cite{Deuretzbacher2008, Deuretzbacher2014, Volosniev2014, Yang2015, Yang2016, Deuretzbacher2016}, and low-energy excitations in the Luttinger liquid model.\ Meanwhile a spinful quantum system in the spin-incoherent regime \cite{Fiete2007} also provides a new avenue for studying 1D quantum many-body systems.\ This regime is termed as spin-incoherent Luttinger liquid (SILL) which forms a different universality class from the Luttinger liquid, where the temperature is high enough that different spin configurations can be regarded as degenerate while low enough that charge excitation is suppressed.\ For 1D spin-$1$ Bose gas \cite{Ho1998, Olshanii1998} with $s$-wave scattering lengths satisfying $|a_0-a_2|$ $\ll$ $a_{0,2}$, there exists a window of temperature for the gas in SILL regime.\ This happens since the sound velocity is much larger than the spin velocity.\

In the crossover regime between Luttinger liquid and SILL, 1D fermions with tunable spins \cite{Pagano2014} and their high momentum tails \cite{Decamp2016} have been studied, which show an evident broadening in the momentum distributions \cite{Cheianov2005, Feiguin2010}.\ Quantum criticality \cite{Hazzard2013} and Pomeranchuk effect \cite{Zhou2014} in the spin-incoherent regime are also theoretically predicted in the two-dimensional Hubbard model.\ Here in contrast we investigate 1D spin-1 Bose gas \cite{Cazalilla2011, Stamperkurn2013} in the SILL regime in a harmonic trap, which is studied only recently \cite{Jen2016_spin1}.\ We shall focus on the Tonks-Girardeau (TG) regime \cite{Tonks1936, Girardeau1960} where the density is sufficiently low that the effective repulsion between particles can be regarded as infinite.\ TG spinor Bose gas is a special case of SILL since the exchange energy vanishes in this limit \cite{Volosniev2014, Yang2015}.\ Therefore TG gas automatically is in the regime of SILL.\ In TG gas limit, we can write down the exact spatial wave functions since bosons are fermionized and impenetrable due to effectively infinitely-strong atom-atom interactions.\ We then numerically calculate the momentum distributions for the three individual components of the spin-1 Bose gas (spin-plus, minus, and $0$).\ These predictions can be measurable in spin-resolved matter-wave experiments, either the time-of-flight experiment \cite{Shvarchuck2002, Davis2012, Jacqmin2012, Fang2016} or Bragg scattering spectroscopy \cite{Kozuma1999, Stenger1999, Steinhauer2002, Papp2008, Veeravalli2008, Pino2011}.\ This system allows for better demonstrations of SILL physics which is within reach of present experimental conditions.\ As compared with electronic spin-$1/2$ systems \cite{Fiete2007, Feiguin2010}, ultracold atom experiments not only provide controllable spatial dimensions but also tunable atom-atom interactions via Feshbach resonances, thus making our investigations testable in the quantum many-body systems.\   

In Ref. \cite{Jen2016_spin1}, we have derived the wave functions and density matrix for 1D spin-$1$ Bose gas in TG limit.\ We numerically calculate its momentum distributions, summed over spin components, up to six bosons.\ The momentum distributions are uniformly broadened as the number of bosons $N$ grows.\ We have also derived the analytical large momentum ($p$) asymptotic in one-body momentum distributions, which shows the universal $1/p^4$ dependence.\ The coefficients of the asymptotic $1/p^4$ are also formulated for arbitrary $N$.\ Here we present the spin-dependent properties of density matrix in 1D spin-$1$ Bose gas in TG limit, and show the spin-dependent momentum distributions up to $N$ $=$ $16$.\ We also obtain the spin-dependent coefficients of $1/p^4$ for asymptotic large $p$.\ Though the momentum distributions vary between different spin components for small $N$, they approach each other as $N$ increases.\ We show this from the numerical results accompanied by analytical arguments in the large $N$ limit.

The rest of the paper is organized as follows.\ In Sec. II we introduce the general wave functions for 1D spin-1 Bose gas.\ In Sec. III, we derive the general forms of density matrices for each spin components with individual spin function overlaps in SILL regime, and present the numerically calculated results using Monte Carlo integration method implemented with Gaussian unitary ensemble.\ In Sec. IV. we discuss the analytical derivation of high momentum asymptotic for each component, which we compare with numerically calculated momentum distributions.\ We also investigate the momentum distributions in large $N$ limit using the method of steepest descent or stationery phase, and compare them with the numerical results.\ Finally we conclude in Sec. V.

\section{General wave functions in TG limit}

In general we can express the wave function of $N$ bosons as
\bea
|\Psi\rangle=\sum_{s_1,s_2,...s_N}\psi_{s_1,s_2,...s_N}(\vec{x})|s_1,s_2,...,s_N\rangle,
\eea
where we denote $\vec{x}$ $=$ $(x_1,x_2,...,x_N)$ and $|s_1,s_2,...s_N\rangle$ $\equiv$ $|\vec s\rangle$ as the spatial distributions and the spin configurations respectively.\ Here we can label $s_i$ $=$ $+$, $-$, or $0$, respectively for spin-plus, minus, and zero components for the $i$th particle.\ The total wave function must satisfy the bosonic symmetry, therefore it is sufficient to just consider the ordered region of $x_1$ $<$ $x_2$ $<$ $...$ $<$ $x_N$, and we can obtain all other regions via permutations of this ordered region.\ In TG gas limit, the atoms become fermionized that their spatial wave functions take the Slater determinant form of noninteracting fermions.\ For the symmetrized spatial part of the wave function, we denote it as $\psi_{\vec{n}}^{sym}(\vec{x})$ which can be expressed in terms of the eigenfunctions $\phi_{n_j}(x_j)$ of the noninteracting fermions in a harmonic trap,
\bea
\psi_{\vec{n}}^{sym}(\vec{x})&=&\frac{1}{\sqrt{N!}}\mathbb{A}[\phi_{n_1}(x_1),\phi_{n_2}(x_2),...,\phi_{n_N}(x_N)]\nonumber\\
&\times&{\rm sgn}(x_2-x_1)\times{\rm sgn}(x_3-x_2) ...\nonumber\\
&\times&{\rm sgn}(x_N-x_{N-1}).\label{wf}
\eea
We denote the sign function as sgn and the anti-symmetrizer as $\mathbb{A}$ for later convenience.\ The orbital indices are ($n_1$, $n_2$, $...n_N$), and the prefactor $\sqrt{N!}$ normalizes the wave function.\ For convenience we use the dimensionless forms of the eigenfunctions $\phi_n(y)$,
\bea
\phi_n(y)&=&\frac{1}{\sqrt{2^n n!}}\frac{1}{\pi^{1/4}}H_n(y)e^{-y^2/2},~y\equiv x/x_{ho},
\eea
where $H_n$ are Hermite polynomials.\ The harmonic oscillator length is $x_{ho}$ $\equiv$ $\sqrt{\hbar/(M\omega)}$ where $\omega$ is the trap frequency and $M$ is the atomic mass.\

To eventually evaluate the density matrix for say the "$+$" component, we need to obtain the wave function amplitude where at least one particle has spin "$+$".\ First we consider some degenerate and normalized spin configuration state $|\chi\rangle$ in some sector of magnetization, and the wave function can be expressed as $|\Psi\rangle$ $=$ $\psi_{\vec{n}}^{sym}(\vec{x})|\chi\rangle$.\ Take $N$ $=$ $3$ for an example, we obtain the probability amplitude $\psi_{+,s_2,s_3}^{sym}(x,x_2,x_3)$ for the first particle having spin "$+$" when we project $|\Psi\rangle$ to $\langle s_1=+,x_1=x|$ in the ordered region of $x$ $<$ $x_2$ $<$ $x_3$.\ To access the probability amplitudes in the other regions, we use the permutation operators $P_{12}$ and $P_{123}$ on the projected states, obtaining
\bea
&&x<x_2<x_3,~\langle (+,s_2,s_3)|\chi\rangle, \nonumber\\
&&x_2<x<x_3,~\langle (s_2,+,s_3)|\chi\rangle=\langle P_{12} (+,s_2,s_3)|\chi\rangle, \nonumber\\
&&x_2<x_3<x,~\langle (s_2,s_3,+)|\chi\rangle=\langle P_{123} (+,s_2,s_3)|\chi\rangle,
\eea
where we have suppressed the common $\psi_{\vec{n}}^{sym}(\vec{x})$ factors.\ Similar construction applies for other $N$'s.\ In the next section we proceed to calculate the spin-dependent density matrices for spin-1 Bose gas in the SILL regime.

\section{Density matrices for SILL of spin-1 Bose gas}

The spin-dependent single-particle density matrix can be straightforwardly written down from the wave function described above.\ For example of the spin-plus component, we have
\bea
\rho_+(x,x')=N\sum_{\vec{s}'}\int d\bar{x}\psi_{+,\vec{s}'}^*(x,\bar{x})\psi_{+,\vec{s}'}(x',\bar{x}),\label{rhop}
\eea 
where $\bar{x}$ $\equiv$ $(x_2,x_3,...,x_N)$ and $\vec s'$ $\equiv$ $(s_2,s_3,...,s_N)$.\ A factor of $N$ represents $N$ possible choices of $x$ and $x'$.\ 

Again we take $N$ $=$ $3$ as an example, and consider only the region of $x$ $<$ $x'$ which is symmetric to $x$ $>$ $x'$.\ The spin-plus single-particle density matrix for $N$ $=$ $3$ then becomes
\begin{widetext}
\bea
\rho_+(x<x')=&& 3\times 2\times\left\{\int_{x<x'<x_2<x_3}(E,E)_+\right.+\int_{x<x_2<x'<x_3}(E,P_{12})_++\int_{x<x_2<x_3<x'}(E,P_{123})_+\nonumber\\
&&+\int_{x_2<x<x'<x_3}(P_{12},P_{12})_++\left.\int_{x_2<x<x_3<x'}(P_{12},P_{123})_++\int_{x_2<x_3<x<x'}(P_{123},P_{123})_+\right\}\nonumber\\
&&\times\psi_{\vec{n}}^{sym*}(x,x_2,x_3)\psi_{\vec{n}}^{sym}(x',x_2,x_3)dx_2dx_3,\label{N3}
\eea
\end{widetext}
where the parentheses $()$ in various integral regions represent the spin function overlaps.\ Take $(E,P_{12})_+$ as an example where $E$ is the identical permutation operator, we define 
\bea
(E,P_{12})_+=\sum_{s_2,s_3}\langle E (+,s_2,s_3)|\chi\rangle \langle P_{12}(+,s_2,s_3)|\chi\rangle.
\eea
Similar forms apply to the other spin function overlaps in Eq. (\ref{N3}).\ Also the factor of $2$ in Eq. (\ref{N3}) comes from the contribution of the integral region $x_2$ $>$ $x_3$, where the spin function overlaps are the same as those with $x_2$ $<$ $x_3$.\

In the SILL regime, we average the above individual spin function overlaps by the total number of spin state configurations, which is denoted as Tr$_\chi (E)$ $\equiv$ $\sum_\chi$ $\langle\chi|E|\chi\rangle$.\ It is simply the trace (Tr) of the identical operator over all spin configurations $|\chi\rangle$ since $\langle\chi|E|\chi\rangle$ $=$ $1$.\ We define the normalized spin function overlap in general as (using the same notation as the non-normalized one for simplicity)
\bea
(P_{12...j},P_{12...k})_+=\frac{\sum_{\vec s'}\langle P_{12...j}(+,\vec s')|P_{12...k}(+,\vec s')\rangle}{\rm{Tr}_\chi(E)},\label{PP}
\eea
where $P_{12...j}$ are $j$-particle permutation operators in the symmetric group $S_N$.\ To derive Eq. (\ref{PP}), we have used the identity $\sum_\chi$ $|\chi\rangle\langle\chi|$ $=$ $1$.\ $(P_{12...j},P_{12...k})_+$ in general represents the spin function overlap from the integration region where the particle at $x$ permutes to just behind $x_j$ while the particle at $x'$ permutes to just behind $x_k$.

In general for arbitrary $N$, we obtain the spin-plus density matrix as
\begin{widetext}
\bea
\rho_+(x<x')=&&N!\bigg\{\int_{x<x'<x_2...<x_N}(E,E)_++\int_{x<x_2<x'...<x_N}(E,P_{12})_+ +\int_{x<x_2<x_3<x'...<x_N}(E,P_{123})_++...\nonumber\\
&&+\int_{x_2<x<x'...<x_N}(E,E)_+ +\int_{x_2<x<x_3<x'...<x_N}(E,P_{12})_++...+\int_{x_2<x_3...<x_N<x<x'}(E,E)_+\bigg\}\nonumber\\
&&\times\psi_{\vec{n}}^{sym*}(x,\bar{x})\psi_{\vec{n}}^{sym}(x',\bar{x})d\bar{x},\label{main_eq}
\eea
\end{widetext}
where we have used the properties of $(P_{12...j},P_{12...j})_+$ $=$ $(E,E)_+$ and $(E,P_{12...j})_+$ $=$ $(P_{12...m},P_{12...m+j-1})_+$ for $j,m$ $\geq$ $2$.\ The first property can be proved from Eq. (\ref{PP}) by using $P_{12...j}^{-1}P_{12...j}$ $=$ $E^{-1}E$ $=$ $1$.\ To prove the second property, we can reduce $P_{12...m}^{-1}P_{12...m+j-1}$ to 
\bea
&&(P_{m-1,m}...P_{23}P_{12})^{-1}P_{m...m+j-1}(P_{m-1,m}...P_{23}P_{12})\nonumber\\
&&=P_{12}^{-1}P_{23}^{-1}...P_{m-1,m}^{-1}P_{m...m+j-1}P_{m-1,m}...P_{23}P_{12},\nonumber\\
&&=P_{1,m+1...m+j-1},\nonumber
\eea
such that $(E,P_{1,m+1...m+j-1})_+$, using again Eq. (\ref{PP}), is exactly the same as $(E,P_{12...j})_+$.

Other spin components of the density matrices, $\rho_-(x<x')$ and $\rho_0(x<x')$, can be derived similarly from characterizing the respective normalized spin function overlaps $()_{-,0}$ which we will evaluate below.\

From now on we limit ourselves to the specific sector of total $S_z$ $\equiv$ $\sum_{i=1}^N s_i$ $=$ $0$.\ For $S_z$ close to $N$, spin-$1$ Bose gas will behave not much different from the polarized or spinless one.\ Therefore we choose the sector of zero $S_z$, which allows the SILL of spin-$1$ Bose gas to distinguish most significantly from the spinless bosons for $S_z$ $\lesssim$ $N$.\ The spin configurations $|\chi\rangle$ in this sector generally involve $n$ pairs of $(+-)$, that is $|+++---00...0\rangle$ with $n$ $=$ $3$ for example.\ The total number of states can then be calculated as
\bea
w_{N}\equiv{\rm Tr}_{\chi}(E)=\sum_{n=0}^{\frac{N}{2}~{\rm or}~\frac{N-1}{2}} \frac{N!}{(n!)^2(N-2n)!},
\eea
which we obtain by permuting $n$ $(\pm)$'s and $(N-2n)$ $(0)$'s.\ For the spin-plus component of the single-particle density matrix in Eq. (\ref{main_eq}), the spin configuration $|00...0\rangle$ with $n$ $=$ $0$ never contributes.\ Therefore we consider only the spin configurations of at least one pair of $(+-)$, and $|\chi\rangle$ can be generally expressed as
\bea
|+\underbrace{+...+}_{n-1}\underbrace{-...-}_{n}\underbrace{00...0}_{N-2n}\rangle.\nonumber
\eea
The first $+$ is projected out in $\rho_+(x<x')$, and thus we have the normalized spin function overlap $(E,E)_+$,
\bea
(E,E)_+&=&\frac{1}{w_N}\sum_{n=1}^{\frac{N}{2}~{\rm or}~\frac{N-1}{2}}\frac{(N-1)!}{(n-1)!n!(N-2n)!},\label{EE_p}
\eea
which is averaged by $w_N$, the total number of states.\ We note that all the arguments of the factorials should be equal and larger than zero.\ $(E,E)_+$ is proportional to the number of states obtained by permuting the rest of $(n-1)$ $(+)$'s, $n$ $(-)$'s, and $(N-2n)$ $(0)$'s. For $(E,P_{12...j})_+$, it has a contribution only when the first $j$ entries are $(+)$'s,
\bea
|\underbrace{+...+}_{j}\underbrace{+...+}_{n-j}\underbrace{-...-}_{n}\underbrace{00...0}_{N-2n}\rangle,\nonumber
\eea
such that we have 
\bea
(E,P_{12...j})_+&=&\frac{1}{w_N}\sum_{n\geq j}^{\frac{N}{2}~{\rm or}~\frac{N-1}{2}}\frac{(N-j)!}{(N-2n)!n!(n-j)!},\label{EP_p}
\eea
which denotes the number of states obtained by permuting the rest of $(n-j)$ $(+)$'s, $n$ $(-)$'s, and $(N-2n)$ $(0)$'s.\ In this specific sector of zero $S_z$, we note that in general $(E,P_{12...j})_+$ is nonvanishing only when $j$ $\leq$ $N/2$.

These corresponding spin function overlaps in $\rho_-(x<x')$, which are $(E,E)_-$ and $(E, P_{12...j})_-$, should be the same as those in $\rho_+(x<x')$.\ While for $\rho_0(x<x')$, we have the spin function overlaps as
\bea
(E,E)_0&=&\frac{1}{w_N}\sum_{n=0}^{\frac{N}{2}~{\rm or}~\frac{N-1}{2}}\frac{(N-1)!}{(n!)^2(N-2n-1)!},\label{EE_0}\\
(E,P_{12...j})_0&=&\frac{1}{w_N}\sum_{n=0}^{\frac{N}{2}~{\rm or}~\frac{N-1}{2}}\frac{(N-j)!}{(n!)^2(N-2n-j)!},\label{EP_0}
\eea
which respectively denote the number of states contributed from the spin configurations with the first one and the first $j$ entries of $(0)$'s.\ We note of the identity that
\bea
2(E,E)_+ + (E,E)_0 = 1.
\eea
This also corresponds to the particle number conservation, that is $2N_+$ $+$ $N_0$ $=$ $N$, where $N_{\pm(0)}$ $\equiv$ $\int dx\rho_{\pm(0)}(x,x)$.\ Thus the number of particles is proportional to the spin function overlaps, $N_{\pm(0)}$ $=$ $N(E,E)_{\pm(0)}$.\ Furthermore we note that
\bea
2 (E,P_{12...j})_+ + (E,P_{12...j})_0 = w_{jN}/w_{N},
\eea
where $w_{jN}$ was defined in Ref. \cite{Jen2016_spin1},
\bea
w_{jN}&\equiv&\sum_{n=0}^{\frac{N}{2}~{\rm or}~\frac{N-1}{2}}\left[\frac{(N-j)!}{(n!)^2(N-2n-j)!}\right.\nonumber\\
&+&\left.\frac{2(N-j)!}{(n-j)!n!(N-2n)!}\right].\label{wjN}
\eea

\begin{figure}[t]
\centering
\includegraphics[width=8.5cm,height=4.2cm]{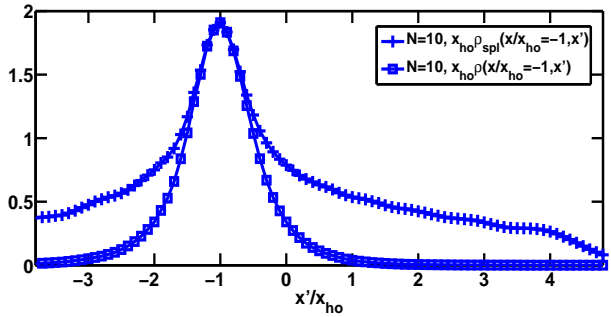}
\caption{(Color online) Single-particle density matrix of spin-1 and spinless bosons for $N$ $=$ $10$ in the sector of $S_z$ $=$ $0$.\ The spatial correlations are plotted at chosen $x/x_{ho}$ $=$ $-1$.\ Spinless (spl) bosons ($+$) show broader spatial distributions than the spin-1 bosons ($\square$), indicating a narrower distribution in momentum space.}\label{fig1}
\end{figure}
\subsection{Spatial correlation in SILL of spin-1 Bose gas}

The effect of the spin function overlaps in the SILL regime can be seen in Fig. \ref{fig1} where we compare the spatial correlations of spin-$1$ [$\rho(x,x')$] and spinless bosons [$\rho_{\rm spl}(x,x')$] at some chosen $x$ in a harmonic trap.\ Spinless or spin-polarized bosons show a wider spatial distribution than the spin-$1$ case, indicating a sharper momentum distribution.\ Large $|x-x'|$ in the spatial correlation corresponds to the small $p$ region.\ For spinless bosons, it has been shown in the bulk that $\rho_{\rm spl, b}(x,x')$ $\propto$ $|x-x'|^{-1/2}$, thus small $p$ behavior is proportional to $|p|^{-1/2}$ \cite{Lenard1964, Vaidya1979, Jimbo1980, Forrester2003}.\ This narrow momentum distribution resembles the one of a Bose-Einstein condensate but not quite since no condensation is allowed \cite{Dalfovo1999} due to large quantum fluctuations in 1D system.\ Therefore no off-diagonal long-range order can be present in the density matrix of 1D Bose gas.\ However a superfluid phase can exist in 1D quantum systems, possessing a power-law decay in spatial correlations.\ This power-law decay can be well described in the Luttinger liquid model using the bosonization method \cite{Giamarchi2004, Haldane1981}. 

In a harmonic trap as shown in Fig. \ref{fig1}, the spatial correlations of $\rho_{\rm spl}(x,x')$ are similar to the one in a bulk in a moderate region of $|x-x'|$ until the correlation decays faster at the edge of the trap ($x'$ $\gtrsim$ $4x_{\rm ho}$).\ In the trap, $\rho_{\rm spl}(p=0)$ is finite.\ It has also been shown that $\rho_{\rm spl}(p=0)$ $\propto$ $N$ in the large $N$ limit \cite{Papenbrock2003}.

In sharp contrast to the spinless bosons in Fig. \ref{fig1}, the spin-$1$ Bose gas in the SILL regime shows an exponential decay in its spatial correlation, which is therefore not condensed.\ This exponential decay has been predicted in the single-particle Green's function of quantum wires in the SILL regime \cite{Cheianov2004, Fiete2004, Fiete2007}, distinguishing from the Luttinger liquid with only power-law decays.\ In the momentum distributions on the other hand, spin-incoherence tends to broaden the distributions, which has been investigated in the $t$-$J$ model \cite{Feiguin2010, Penc1996, Penc1997} or the system of uniform two-component gas \cite{Cheianov2005}.\ Similarly the spin-$1$ bosons in the SILL regime will also have a broadened momentum distribution due to the averaging of the spin function overlaps, which we discuss in more details below.\ Large $p$ behavior will be discussed later in Sec. IV. A.

\subsection{Momentum distribution in SILL of spin-1 Bose gas}

\begin{figure*}[t]
\centering
\includegraphics[width=13.5cm,height=6.2cm]{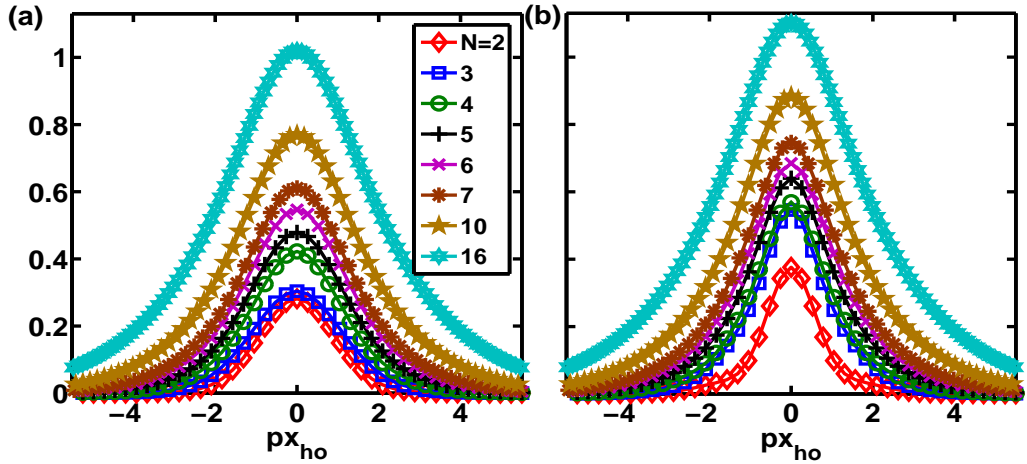}
\caption{(Color online) Momentum distributions of (a) spin-plus [$\rho_+(p)$] and (b) zero components [$\rho_0(p)$] in the SILL regime.\ In the sector of $S_z$ $=$ $0$, we numerically calculate the momentum distributions of 1D spin-1 TG gas up to $N$ $=$ $16$.\ They are uniformly broadened as $N$ increases, and the peaks of $\rho_0(p)$ are larger than $\rho_+(p)$ due to the spin function overlaps.}\label{fig2}
\end{figure*}

We define the spin-dependent momentum distributions as 
\bea
\rho_{\pm(0)}(p)=\frac{1}{2\pi}\int_{-\infty}^\infty dx \int_{-\infty}^\infty dx' e^{ip(x-x')}\rho_{\pm(0)}(x,x'),\label{momentum}
\eea
where we set $\hbar$ $=$ $1$.\ We then numerically calculate the momentum distributions of the three components in 1D TG Bose gas based on Eq. (\ref{main_eq}), $\rho_-(x<x')$, and $\rho_0(x<x')$.\ In Fig. \ref{fig2}, both spin components of $\rho_+(p)$ and $\rho_0(p)$ are uniformly broadened as $N$ grows, and $\rho_+(p)$ $\neq$ $\rho_0(p)$ for finite $N$.\ The effect of spin-incoherence also averages out the oscillatory structure that is present in the momentum distribution for specific spin state of spinor Bose gas \cite{Deuretzbacher2008}.\ Furthermore the peaks of $\rho_0(p)$ are larger than $\rho_+(p)$ up to $N$ $=$ $16$.\ This is due to $N_0$ $\geq$ $N_\pm$ in general and the spin function overlaps $(E,P_{12...j})_0$ are always larger than $(E,P_{12...j})_+$, which we will show more specifically in Fig. \ref{fig9} in Appendix.\ For spinless bosons, the peaks of $\rho_{\rm spl}(p)$ have a scaling of $\rho_{\rm spl}(p=0)$ $\propto$ $N$ \cite{Papenbrock2003}.\ Here the spin-$1$ Bose gas in the SILL regime shows fitted scalings of $\rho_+(p=0)$ $\propto$ $N^{0.49}$ and $\rho_0(p=0)$ $\propto$ $N^{0.66}$ from Fig. \ref{fig2}.\ These reduced scalings again show the feature of broadened momentum distributions in the SILL regime

To calculate $\rho_{\pm(0)}(x,x')$ we implement Gaussian unitary ensemble (GUE) \cite{Papenbrock2003} to speed up the convergence in the Monte Carlo (MC) integration method.\ The GUE draws a series of $(N-1)$ random numbers in $\bar x$, which are repulsively distributed due to the joint probability density of $\Pi_{1\leq i<j\leq N-1}(x_i-x_j)^2$.\ This implementation of GUE thus enables our MC integration to simulate up to $N$ $=$ $16$, which in this case takes about $140$ hours with MC simulations of $M$ $=$ $10^6$ sets of random numbers using $200$ parallel CPU cores.\ All MC simulations in Fig. \ref{fig2} use $M$ $=$ $10^7$ except for $N$ $=$ $16$ with $M$ $=$ $10^6$.

In the next section we investigate their asymptotic forms in large momentum limit, which show $1/p^4$ decay, and their momentum distributions in large $N$ limit.

\section{Momentum distributions in high \texorpdfstring{$\boldsymbol{p}$}{p} and large \texorpdfstring{$\boldsymbol{N}$}{N} limits} \label{largeN}

\subsection{Asymptotic high \texorpdfstring{$\boldsymbol{p}$}{p} limit}

For spinless bosons in the TG limit, relative wave function between two particles in short distance is $\psi_{\rm rel}(x,x')$ $\propto$ $|x-x'|$, indicating of impenetrable bosons and corresponding to the feature of fermionic repulsion.\ Again it has been shown \cite{Lenard1964, Vaidya1979, Jimbo1980, Forrester2003} in a bulk where $\rho_{\rm spl, b}(x,x')$ in short distance is proportional to $[1+...+|x-x'|^3/(9\pi)+...]$.\ Thus the non-analytic $|x-x'|^3$ term in the short-distance correlation gives a universal $1/p^4$ asymptotic in large momentum limit.\ This universal $1/p^4$ asymptotic is not unique for a Bose gas with two-body contact interactions \cite{Minguzzi2002, Olshanii2003, Xu2015}.\ It also shows up in Tan's relation \cite{Tan2008, Barth2011} in the two-component Fermi gas \cite{Braaten2008-1, Braaten2008-2,Werner2009, Zhang2009}.\ 

For 1D spin-$1$ TG Bose gas on the other hand, the analytical results for a high $p$ asymptotic total momentum distribution $\rho(p)$ have been derived \cite{Jen2016_spin1}, showing also a universal $1/p^4$ dependence.\ Similarly for its spin-dependent components, they can be straightforwardly written as 
\bea
\rho_{\pm(0)}(p)\underset{p\rightarrow\infty}{=}&&\frac{2[(E,E)_{\pm(0)}+(E,P_{12})_{\pm(0)}]}{2\pi p^4}\nonumber\\
&&\times\sum_{(n_i,n_j)}\int_{-\infty}^\infty dx
\left| \begin{array}{cc}
\phi_{n_i}'(x) & \phi_{n_j}'(x) \\
\phi_{n_i}(x) & \phi_{n_j}(x) \end{array} \right|^2,\label{large_p}
\eea 
where $(n_i,n_j)$ denotes any possible pairs of $N$ harmonic oscillator eigenfunctions.\ The asymptotic form depends on the spin function overlap $(E,P_{12})_{\pm(0)}$ because it has significant contributions only from the integral regions of $x$ $<$ $x_j$ $<$ $x'$ and $x'$ $<$ $x_j$ $<$ $x$ for all $x_j$ $\in$ $\bar{x}$ with $x$ $\approx$ $x'$.\ The asymptotic form for the spinless bosons can be also obtained by replacing $[(E,E)_{\pm(0)}+(E,P_{12})_{\pm(0)}]$ with $2$ in Eq. (\ref{large_p}).\ We note that using Eqs. (\ref{large_p}) and (\ref{wjN}), we have $2\rho_+(p)$ $+$ $\rho_0(p)$ $=$ $\rho(p)$ where the last quantity was computed in Ref. \cite{Jen2016_spin1}.\

\begin{figure}[b]
\centering
\includegraphics[width=8.5cm,height=4.5cm]{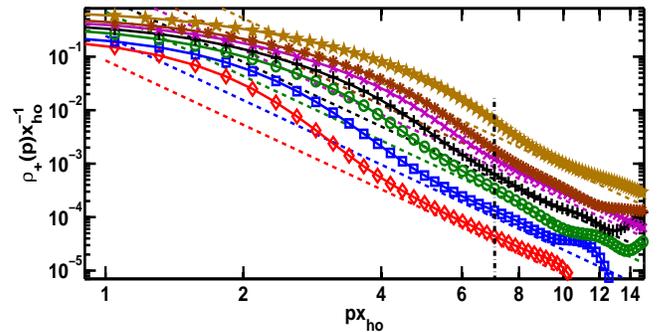}
\caption{(Color online) Asymptotics of high momentum distributions in Fig. \ref{fig2}.\ High momentum distributions are plotted in logarithmic scales and compared with analytic calculations (dash) in the spin-plus component of 1D spin-1 TG Bose gas.\ The analytical asymptotics are $(0.085$, $0.245$, $0.694$, $1.267$, $2.131$, $3.225$, $8.305)/p^4$, respectively for $N$ $=$ $2-7$ and $10$.\ The line symbols and colors are the same as Fig. \ref{fig2}, and a vertical line (dash-dot) at around $px_{ho}$ $\sim$ $7$ guides the eye for the limitation in accuracy of the numerical calculations.}\label{fig3}
\end{figure}

The spin-$1$ Bose gas in the SILL regime shows very different properties from the spin-coherent ones in the coefficients of high $p$ asymptotics.\ The coefficients are always less than the ones in spinless bosons since $(E,E)_{\pm(0)}$, $(E,P_{12})_{\pm(0)}$ $<$ $1$.\ And for large $N$, $[(E,E)_{\pm(0)}+(E,P_{12})_{\pm(0)}]$ $\rightarrow$ $[1/3+(1/3)^2]$ $=$ $4/9$ from Eq. (\ref{largeN2}), less than $2$ for the case of spinless bosons as well.\ As an example, in Fig. \ref{fig3} we compare the numerical and analytical results of $\rho_+(p)$ in high $p$ limit.\ The numerically calculated high $p$ asymptotics approach approximately to the analytical ones.\ For even larger $px_{ho}$ $\gtrsim$ $7$, the trends either drop and cross the analytical asymptotics, or bounce back and oscillate, indicating the inaccuracy of numerical results in these regions.\ To reach accurate high $p$ asymptotics is quite demanding in MC integrations and consuming more CPU time for even larger $N$.\ However, MC simulations have already achieve the accuracy of $10^{-3}$ and $10^{-2}$ of the momentum distributions for $N$ $=$ $2-3$ and $10$ respectively.

We have also evaluated numerically the potential $(\langle V\rangle)$ and kinetic energies $(\langle K\rangle)$.\ Since our 1D bosonic TG gas has the same density distribution as the one of a Fermi gas, we have $\langle V\rangle$ $=$ $\langle K\rangle$ $=$ $N^2\hbar\omega/4$, equivalent to half of the total energy, which complies with the Virial theorem \cite{Werner2006, Werner2008}.\ In Ref. \cite{Jen2016_spin1}, we concatenate $\rho(p)$ with the asymptotic tails analytically derived to improve the energy calculations.\ Here we directly use the momentum distributions of $\rho(p)$ calculated by MC simulations implemented with GUE.\ We find that the numerical results of these energies improve to the relative errors below $7\%$ and $10\%$ for $N$ $=$ $2-7$ and $16$ respectively to the exact values of $\langle V\rangle$ and $\langle K\rangle$.\ This further shows the advantage of GUE in the convergence and accuracy of our numerical results.

\begin{figure}[t]
\centering
\includegraphics[width=8.5cm,height=4.5cm]{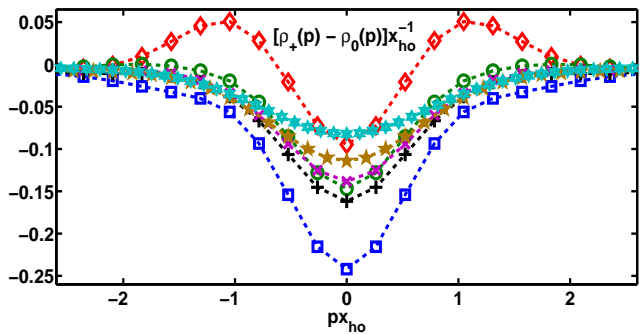}
\caption{(Color online) The difference between spin-plus and spin-$0$ momentum distributions.\ The symbols and colors are plotted correspondingly as in Fig. \ref{fig2}, where we choose $N$ $=$ $2-6$ and $16$.\ The case of $N$ $=$ $7$ is neglected here for it is too close to the one of $N$ $=$ $6$.\ The difference diminishes gradually as $N$ increases.}\label{fig4}
\end{figure}

In Fig. \ref{fig4}, we plot the difference of spin-plus and zero momentum distributions numerically, $\rho_+(p)$ $-$ $\rho_0(p)$, from Fig. \ref{fig2}.\ The difference goes away gradually as $N$ increases, indicating these two components approach each other in large $N$ limit.\ The dips at around $p$ $\sim$ $0$ demonstrate that the peaks of $\rho_0(p)$ are always higher than $\rho_+(p)$, which is due to larger spin function overlaps for the spin-$0$ component.\ A special feature of the peaks for the case of $N$ $=$ $2$ shows a wider $\rho_+(p)$ than $\rho_0(p)$ while this feature is not obvious for larger $N$.

\subsection{Large \texorpdfstring{$\boldsymbol{N}$}{N} limit}

Due to the limits of numerical integration, we can only calculate the single-particle density matrix of spin-$1$ Bosons up to $N$ $=$ $16$.\ For finite $N$, we have demonstrated numerically $\rho_+(p)$ $\neq$ $\rho_0(p)$ since in general $N_0$ $\geq$ $N_+$ and spin function overlaps $(E,P_{12...j})_0$ are always larger than $(E,P_{12...j})_+$.\ Thus the peaks of $\rho_0(0=0)$ are larger than $\rho_+(p=0)$.\ In this subsection we attempt to investigate the momentum distributions in large $N$ limit.\ The study in this limit can give insight to practical experiments where several hundreds or thousands of atoms are involved.\ 

To investigate the individual components of spin-$1$ momentum distributions in large $N$ limit, we need the asymptotic forms for various spin function overlaps.\ These spin function overlaps in general can be written as
\bea
(E,E)_+&=&\frac{f^{(N-1)}_1}{f^{(N)}_0},~ (E,P_{12...j})_+=\frac{f^{(N-j)}_j}{f^{(N)}_0},\nonumber\\
(E,E)_0&=&\frac{f^{(N-1)}_0}{f^{(N)}_0},~ (E,P_{12...j})_0=\frac{f^{(N-j)}_0}{f^{(N)}_0},\label{relation}
\eea
where 
\bea
f^{(N)}_k\equiv \sum_{j=0}^{\frac{N-k}{2}}\frac{N!}{(k+j)!j!(N-2j-k)!}.
\eea
$f^{(N)}_k$ are just the coefficients of $x^k$ in the binomial expansions of $(x+x^{-1}+1)^N$.\ We then further express them in terms of the complex integral as shown in Appendix, and find the asymptotic form of $f^{(N)}_k$ in large $N$ limit using the method of steepest descent or stationary phase \cite{Bender1999}.\ In Fig. \ref{fig9} of Appendix, the asymptotic forms in Eq. (\ref{fNk}) are used to compare with the exact ones of Eqs. (\ref{EE_p}), (\ref{EP_p}), (\ref{EE_0}), and (\ref{EP_0}), and they approach to the exact ones in large $N$ limit for small $j$ in Eq. (\ref{relation}).\ Therefore we shall use the asymptotic forms to compare the spin-plus with the spin-$0$ components of the spin function overlaps for large $N$. 

\begin{figure}[b]
\centering 
\includegraphics[width=8.5cm,height=4.5cm]{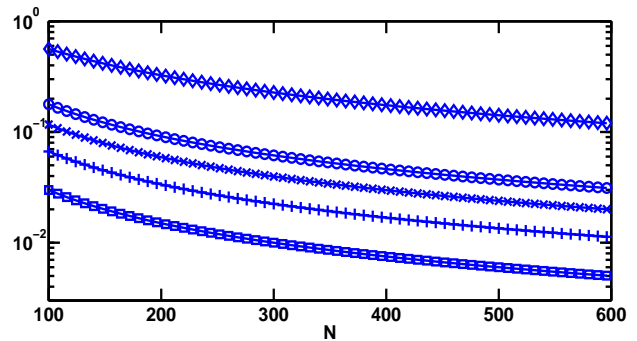}
\caption{(Color online) Relative deviations of Eq. (\ref{deviation}) for the asymptotic forms $\bar f^{(N-j)}_j$ and $\bar f^{(N-j)}_0$ of the exact spin function overlaps $(E,P_{12...j})_+$ and $(E,P_{12...j})_0$ respectively.\ The relative deviations are plotted for $j$ $=$ $2$($\square$), $3$($+$), $4$($\times$), $5$($\circ$), and $10$($\diamond$).\ The deviations decrease as $N$ increases, which indicate that the asymptotic forms are approaching each other.\ It suggests that spin-plus and zero components of the momentum distributions coincide in large $N$ limit.}\label{fig5}
\end{figure}

We define their relative deviations as
\bea
\left|\frac{(E,P_{12...j})_{+}-(E,P_{12...j})_{0}}{(E,P_{12...j})_{0}}\right|,
\eea
which asymptotically approaches to
\bea
\left|\frac{\bar f^{(N-j)}_{j}-\bar f^{(N-j)}_{0}}{\bar f^{(N-j)}_{0}}\right|,\label{deviation}
\eea
in large $N$ limit, where $\bar f^{(N-j)}_j$ is the asymptotic form of $f^{(N-j)}_j$.\ This asymptotic form allows us to compute the deviations for even larger $N$ than using the exact formulas.\ In Fig. \ref{fig5}, the relative deviations decay as $N$ increases for small $j$ and become below $10^{-2}$ for $N$ $\sim$ $600$ with $j$ $\leq$ $3$.\ Note that for a moderate $j$ $=$ $10$, it only reaches $0.1$ for as high as $N$ $=$ $600$.\ Since the spin function overlaps with smaller $j$ contribute much more to the momentum distributions than $j$ $\lesssim$ $N/2$ [see Figs. \ref{fig9}(a) and (b)], along with much less relative deviations, we expect that the spin-plus momentum distribution approaches the one of spin-$0$ as $N$ increases.\ We have also shown this trend in Fig. \ref{fig4} for finite $N$ up to sixteen particles.

\begin{figure}[b]
\centering
\includegraphics[width=8.5cm,height=4.5cm]{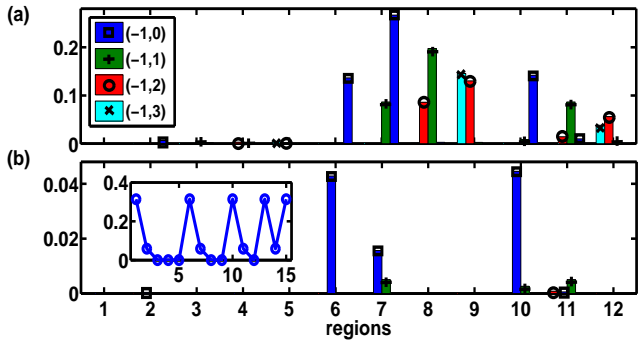}
\caption{(Color online) Contributions from the integral regions for $\rho_+(x<x')$ with $N$ $=$ $5$.\ The values for each integral regions without and with the multiplications of spin function overlaps are shown in (a) and (b) respectively.\ Four spatial correlations are chosen at $(x,x')$ as denoted in the legend of (a) along with the symbols $(\square, +,\circ,\times)$ for clarity.\ The inset in (b) displays $15$ spin function overlaps for the case of five bosons.}\label{fig6}
\end{figure}

This can be further confirmed by studying respective contributions from the integral regions in the single-particle density matrix $\rho_+(x<x')$ of Eq. (\ref{main_eq}).\ In Fig. \ref{fig6} we plot the results of most significant $12$ out of a total $15$ integral regions for the case of $N$ $=$ $5$ with and without the multiplications of spin function overlaps.\ The order of the integral regions can be seen from Eq. (\ref{main_eq}).\ We denote the first five regions as $x$ $<$ $x_2$ with $x'$ $<$ $x_2$ moving sequentially to $x_5$ $<$ $x'$, the next four regions ($6$th to $10$th) as $x_2$ $<$ $x$ $<$ $x_3$ with $x_2$ $<$ $x'$ $<$ $x_3$ moving sequentially to $x_5$ $<$ $x'$, and so on for the rest of the regions.\ A feature of hard core bosons in 1D harmonic trap is that atoms prefer to distribute evenly in space.\ The particles at $\bar x$ repel each other due to their strongly repulsive interactions in TG limit.\ Thus more significant peaks of the integral contributions occur roughly at the corresponding integral regions depending on $(x,x')$.\ For example in Fig. \ref{fig6}(a), the spatial correlation at $(x,x')$ $=$ $(-1,0)$ has the largest contribution at the seventh integral region, $x_2$ $<$ $x$ $<$ $x_3$ $<$ $x'$ $<$ $x_4$ $<$ $x_5$.\ Similarly the spatial correlation at $(x,x')$ $=$ $(-1,3)$ has the largest contribution at the ninth integral region, $x_2$ $<$ $x$ $<$ $x_3$ $<$ $x_4$ $<$ $x_5$ $<$ $x'$.\ In Fig. \ref{fig6}(b) we multiply these values with the spin function overlaps (inset), and show that significant contributions come from the overlaps with small $j$.\ For four specific spatial correlations we choose here, significant values dwell in the $6$th, $7$th, $10$th, and $11$th integral regions, which correspond to the spin function overlaps $(E,E)_+$, $(E,P_{12})_+$, $(E,E)_+$, and $(E,P_{12})_+$ respectively.\ Moreover as expected the spatial correlations decay as $|x-x'|$ increases, reminiscent of the exponential decay discussed in III. A.

For TG bosons in the bulk with a length $L$ in thermodynamic limit, we can use the analytical expression of Eq. (\ref{large_p}) and let $\phi_{n_i}(x)$ $=$ $e^{ik_ix}/\sqrt{L}$ with various eigenmodes $n_i$ $\rightarrow$ $k_i$.\ The spatial integral in Eq. (\ref{large_p}) can then be calculated as 
\bea
&&\int dx \left| \begin{array}{cc}
\phi_{n_i}'(x) & \phi_{n_j}'(x) \\
\phi_{n_i}(x) & \phi_{n_j}(x) \end{array} \right|^2\nonumber\\
&&=\frac{1}{L^2}\int dx
\left| \begin{array}{cc}
ik_ie^{ik_ix} & ik_je^{ik_jx} \\
e^{ik_ix} & e^{ik_jx} \end{array} \right|^2,\nonumber\\
&&=\frac{(k_i-k_j)^2}{L}.
\eea
In the continuous limit of $(k_i,k_j)$ corresponding to the pairs of $(n_i,n_j)$ in the summation, we have $\rho_{\rm spl, b}(p\rightarrow\infty)$ in the bulk
\bea
\frac{\rho_{\rm spl, b}(p\rightarrow\infty)}{L}=\frac{2}{\pi p^4}\int_{-k_F}^{k_F}\frac{dk_i}{2\pi}\int_{-k_F}^{k_i}\frac{dk_j}{2\pi}(k_i-k_j)^2,\nonumber\\
\eea
where $k_F$ $=$ $\pi N/L$.\ Let $k_i$ $=$ $k_Fx$ and $k_j$ $=$ $k_Fy$, the above becomes
\bea
\frac{\rho_{\rm spl, b}(p\rightarrow\infty)}{L}&&=\frac{k_F^4}{2\pi^3p^4}\int_{-1}^1 dx\int_{-1}^x dy(x-y)^2,\nonumber\\
&&=\frac{2k_F^4}{3\pi^3p^4},\label{splb}
\eea
which is the same as Eq. (67) in Ref. \cite{Forrester2003} up to a factor of $2\pi$ in our definition of Fourier transform in Eq. (\ref{momentum}).\ Our derivation is parallel to using short-distance expansions in $\rho_{\rm spl,b}(x,x')$ \cite{Forrester2003} where its non-analytic term $|x-x'|^3/(9\pi)$ after Fourier transformed gives the same result.

For large $N$ spinless bosons in a harmonic trap, we can use the local-density approximation (LDA) for the local chemical potential in Thomas-Fermi limit, which reads $\mu(x)$ $=$ $\mu$ $-$ $\alpha x^2/2$ with $\alpha$ $=$ $M\omega^2$.\ The cutoff momentum can be derived as $k_F(x)$ $=$ $\sqrt{2M(\mu-\alpha x^2/2)}$.\ Since the maximum mode in 1D hard core bosons at infinite interactions is approximately $n_{\rm max}$ $\approx$ $N$, the chemical potential can be determined as $\mu$ $=$ $n_{\rm max}\omega$ $\approx$ $N\omega$.\ We further define the boundary of $x_{\rm max}$ $=$ $\sqrt{2N\omega/\alpha}$ such that we re-express $k_F(x)$ as $\sqrt{(M\alpha)(x_{\rm max}^2-x^2)}$.\ Applying LDA to Eq. (\ref{splb}), we have the density matrix for spinless bosons in a harmonic trap as
\bea
\rho_{\rm spl, LDA}(p\rightarrow\infty)&&=\frac{2}{3\pi^3}\int_{-x_{\rm max}}^{x_{\rm max}}dx\frac{k_F^4(x)}{p^4},\nonumber\\
&&=\frac{2(M\alpha)^2}{3\pi^3p^4}\int_{-x_{\rm max}}^{x_{\rm max}}dx(x_{\rm max}^2-x^2)^2,\nonumber\\
&&=\frac{2^7\sqrt{2}}{45\pi^3}N^{5/2}\frac{1}{p^4x_{ho}^3}, \label{LDA}
\eea
where the scaling of $N^{5/2}$ disagrees with Ref. \cite{Olshanii2003} which gave a different scaling of $N^{3/2}$.\ $N^{5/2}$ scaling has also been reported for the 1D SU($\kappa$) Fermi gas with $\kappa$ $\neq$ $1$ \cite{Decamp2016}.\ The coefficient in Eq. (\ref{LDA}) is the same as the TG Fermi gas in the $\kappa$ $\rightarrow$ $\infty$ limit.\ We note that the coefficient of the scaling depends on the many-body state and is related to the slope of energy ($-dE/dg_{\rm 1D}^{-1}$) \cite{Olshanii2003, Decamp2016}.

For the spinful case of our spin-$1$ Bose gas in large $N$ limit, the asymptotic coefficient of $[(E,E)_{\pm(0)}+(E,P_{12})_{\pm(0)}]$ becomes $[1/3+(1/3)^2]$ $=$ $4/9$ according to Eq. (\ref{largeN2}), such that the spin-dependent and total momentum distributions respectively in thermodynamic limit become
\bea
\rho_{\pm(0)}(p\rightarrow\infty)|_{N\rightarrow \infty}&&=\frac{1}{2}\frac{4}{9}\times\rho_{\rm spl, LDA}(p\rightarrow\infty),\nonumber\\
&&=\frac{2^8\sqrt{2}}{405\pi^3}N^{5/2}\frac{1}{p^4x_{ho}^3},\label{plus}
\eea
and
\bea
\rho(p\rightarrow\infty)|_{N\rightarrow \infty} &&= 3\times\rho_{\pm(0)}(p\rightarrow\infty)|_{N\rightarrow \infty},\nonumber\\
&&=\frac{2^8\sqrt{2}}{135\pi^3}N^{5/2}\frac{1}{p^4x_{ho}^3},
\eea
which is $2/3$ of Eq. (\ref{LDA}).

\begin{figure}[t]
\centering
\includegraphics[width=8.5cm,height=4.5cm]{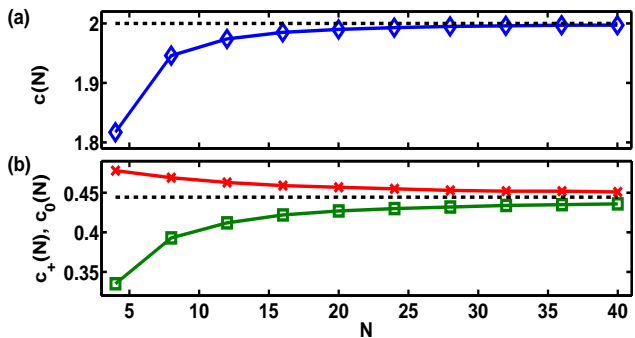}
\caption{(Color online) Comparisons of the coefficients in the momentum distributions of spinless and spin-$1$ Bose gas.\ We plot the coefficients of (a) $c(N)$ ($\diamond$) and (b) $c_{+}(N)$ ($\square$) and $c_0(N)$ ($\times$), respectively for spinless bosons, spin-plus and spin-$0$ components in high $p$ limit.\ These coefficients approach to the asymptotic values $c(\infty)$ $=$ $2$ (dash) and $c_{+(0)}(\infty)$ $=$ $4/9$ (dash) when $N$ increases.}\label{fig7}
\end{figure}

We denote the momentum distribution of spinless bosons in high $p$ limit as $\rho_{\rm spl}(p\rightarrow\infty)$ which can be derived by replacing $[(E,E)_{\pm(0)}+(E,P_{12})_{\pm(0)}]$ with $2$ in Eq. (\ref{large_p}).\ We then define $c(N)$ $\equiv$ $\rho_{\rm spl}(\infty)(2^6\sqrt{2})^{-1}N^{-5/2}(45\pi^3p^4x_{ho}^3)$ and $c(\infty)$ $=$ $2$ according to Eq. (\ref{LDA}).\ $ c_{\pm(0)}(N)$ can be also defined in the same way for $\rho_{\pm(0)}(\infty)$.\ These coefficients can be calculated using Eq. (\ref{large_p}).\ In Fig. \ref{fig7} we plot $c(N)$ and $c_{+(0)}(N)$ respectively to show how they approach the asymptotic large $N$ values.\ We find that Eqs. (\ref{LDA}) and (\ref{plus}) already give good enough estimates for $N$ $\gtrsim$ $20$ and $30$ for spinless and spin-$1$ bosons respectively.\ The relative deviations $|c(N)-c(\infty)|/c(\infty)$ $=$ $0.14\%$ for $N$ $=$ $40$, and for the cases of $c_{+(0)}(40)$, they reach $1.9\%(1.5\%)$.\ In Fig. \ref{fig7}(b), $c_0$ $>$ $c_+$, which again indicates that the spin function overlaps $(E,P_{12...j})_0$ $>$ $(E,P_{12...j})_+$ and $N_0$ $>$ $N_+$.

\section{Conclusion}

In conclusion, we have investigated the spin-dependent properties of spin-$1$ Bose gas in the regime of spin-incoherent Luttinger liquid (SILL).\ Three components (spin-plus, zero, and minus) of the single-particle density matrix for this universal class can be calculated by deriving respective spin function overlaps.\ These spin function overlaps result from the highly degenerate spin configurations in the SILL regime.\ In contrast with the spinless bosons with a power-law decay in its spatial correlation, spin-$1$ bosons in TG limit show an exponentially decaying spatial correlation, which indicates a broadened momentum distribution and a different universal class from Luttinger liquid.\ The universal $1/p^4$ dependence in high $p$ limit is also present in the spin-dependent momentum distributions.\ This asymptotic has a scaling of $N^{5/2}$ with a reduced coefficient than the one of the spinless bosons.\ The coefficients of the asymptotic are proportional to Tan's contact and can be observed in experiments as one of the signatures of SILL.\ We compare these analytical predictions with the numerical results calculated by Monte Carlo (MC) integration with Gaussian unitary ensemble (GUE) up to sixteen bosons.\ The method of MC integration implemented with GUE converges faster and gives more accurate results such that we are able to calculate higher $p$ regions.\ The high momentum tails approximately and asymptotically follow the reduced coefficients we analytically derived.

For the $S_z$ $=$ $0$ sector, we show that the spin-$0$ component always has a larger peak than the spin-plus momentum distribution for finite $N$.\ This can be explained by the spin function overlaps which are larger for the spin-$0$ density matrix than the spin-plus case.\ While they differ for small $N$, they coincide in the large $N$ limit.\ This indicates that highly incoherent bosons form in this limit with each component occupying exactly one third of the total number of particles.\ The ultracold spinor Bose gas allows for a potential realization of this universal class of SILL, and our results offer a testable paradigm to study quantum many-body phenomena in 1D strongly interacting bosons.

\section*{Acknowledgements}
This work is supported by the Ministry of Science and Technology, Taiwan, under Grant No. 104-2112-M-001-006-MY3.\ The work of SKY was partially supported by a grant from the Simons Foundation, and was performed at the Aspen Center for Physics supported by National Science Foundation Grant No. PHY-1066293.
\appendix
\section{Spin function overlaps in large \texorpdfstring{$\boldsymbol{N}$}{N} limit}

Before we derive the spin function overlaps in large $N$ limit, first we express them in terms of an integral function, and then introduce the method of stationary phase or steepest descent to solve for their asymptotic forms \cite{Bender1999}.\ Consider the following function,
\bea
f^{(N)}(x)&=&(x+x^{-1}+1)^N,\nonumber\\
&=&\sum_{j_2=0}^N\sum_{j_1=0}^{N-j_2}\frac{N!}{j_1!j_2!(N-j_1-j_2)!}x^{j_1-j_2},
\eea
where the second line above can be derived by binomial expansions.\ When we set $k$ $=$ $j_1$ $-$ $j_2$ and $j$ $=$ $j_2$, we have
\bea
f^{(N)}(x)=\sum_{k=-N}^N\sum_{j=0}^{\frac{N-k}{2}}\frac{N!}{(k+j)!j!(N-2j-k)!}x^k,\label{f}
\eea
where the upper bound of index $j$ can be derived by solving $j_2$ in the equations of $j_1$ $+$ $j_2$ $=$ $N$ and $j_1$ $-$ $j_2$ $=$ $k$.\ From Eq. (\ref{f}), we define the coefficient of $x^k$ in $f^{(N)}(x)$ as $f^{(N)}_k$ which is the same as the one of $x^{-k}$.\ Compare with Eqs. (\ref{EE_p}), (\ref{EP_p}), (\ref{EE_0}), and (\ref{EP_0}) in the main paper, we find that the spin function overlaps can be expressed as Eq. (\ref{relation}) in the main text.\ These coefficients can be calculated as follows
\bea
f^{(N)}_k=\frac{1}{2\pi i}\oint_C dz z^{k-1}(z+z^{-1}+1)^N,
\eea
where the coefficient $f^{(N)}_k$ (of $z^{-k}$ in this case) is exactly the residue at the pole of $z$ $=$ $0$ with a complex number $z$.\ $C$ in the above denotes a contour integration on a circle in a counterclockwise direction around the origin.

The asymptotic behavior of the integral can be obtained in large $N$ limit.\ We let $z$ $=$ $e^{i\theta}$, the coefficient becomes
\bea
f^{(N)}_k&=&\frac{1}{2\pi}\int_{-\pi}^\pi d\theta e^{ik\theta}(1+2\cos\theta)^N\nonumber\\
&=&\frac{1}{2\pi}\int_{-\pi}^\pi d\theta e^{\mathcal{F}(\theta)},\label{integral}
\eea
where $\mathcal{F}(\theta)$ $\equiv$ $N\ln(1+2\cos\theta)+ik\theta$.\ Let $\theta$ $\rightarrow$ $w$ in the complex plane, using the method of steepest descent \cite{Bender1999} for the above highly-oscillating integrals in large $N$ limit, we first find the saddle points which satisfy the first derivative $\mathcal{F}'(w)$ $=$ $0$.\ The integrals are then dominated by the local maxima passing the saddle points along the integration contour.

\begin{figure}[t]
\centering
\includegraphics[width=8cm,height=5cm]{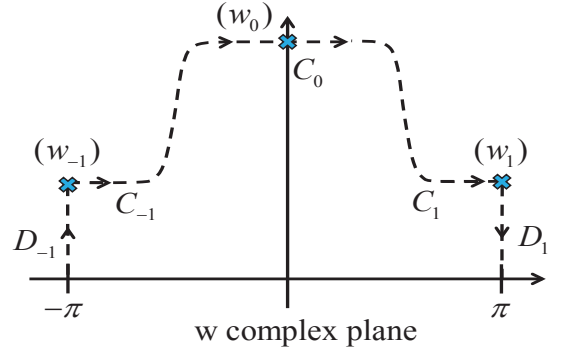}
\caption{(Color online) Integration paths for calculating the asymptotic form of $f^{(N)}_k$ traversing through three saddle points (denoted as $\times$).\ Starting from $w$ $=$ $-\pi$ in the complex plane of $w$, we choose a vertical path $D_{-1}$ to reach the first saddle $w_{-1}$ with increasing imaginary parts only.\ Paths $C_{\pm 1}$ traverse through the saddle points $w_{\pm 1}$ from the left tangent to the real axis, which connect to the path $C_0$ crossing the third one $w_0$, again tangent to the real axis.\ The last path $D_1$ reaches the end point of $w$ $=$ $\pi$ starting from $w_1$.\ Paths $C_{\pm 1}$ and $C_0$ follow the valleys of steepest descent while $D_{\pm 1}$ do not.}\label{fig8}
\end{figure}

The saddle points are therefore located at
\bea
-2\sin w+\frac{ik}{N}(1+2\cos w)=0,
\eea
which, after replacing trigonometry functions with exponentials, becomes
\bea
e^{iw}=\frac{-k/N\pm\sqrt{(k/N)^2+4(1-k^2/N^2)}}{2(1+k/N)}.\label{roots}
\eea
If $k$ $\rightarrow$ $0$, the above suggests a multiple of roots that satisfy $e^{iw}$ $=$ $\pm 1$, which are $w$ $=$ $2n\pi$ and $\pm(2n+1)\pi$ for integers $n$, indicating multiple saddle points in this integral.\ We consider an integration path in Fig. \ref{fig8}, where only three saddle points ($n$ $=$ $0$) are involved.\ Below we demonstrate why the contour is valid and guarantees to follow the valleys of steepest descent between these three saddle points, which can be determined by the sign of $\mathcal{F}''(w)$.

First to calculate $\mathcal{F}''(w_0)$, we define $Q$ $\equiv$ $\sqrt{4-3(k/N)^2}$, and we have from Eq. (\ref{roots}) with the "$+$" sign,
\bea
e^{iw_0}=\frac{-k/N+Q}{2(1+k/N)},~ e^{-iw_0}=\frac{k/N+Q}{2(1-k/N)},
\eea
where $w_0$ should be purely imaginary in general.\ We further use the above to reinterpret 
\bea
\cos w_0&=&\frac{(k/N)^2+Q}{2[1-(k/N)^2]},~ 1+2\cos w_0=\frac{1+Q}{1-(k/N)^2},\nonumber\\
\sin w_0&=&\frac{i}{2}\frac{k}{N}\frac{1+Q}{1-(k/N)^2}.
\eea
Now the second derivative of $\mathcal{F}(w_0)$ becomes
\bea
\mathcal{F}''(w_0)&=&N\left[\frac{-2\cos w_0}{1+2\cos w_0} - \frac{4\sin^2 w_0}{(1+2\cos w_0)^2}\right],\nonumber\\
&=&-N\frac{[Q(1-(k/N)^2)]}{1+Q},\label{eq1}
\eea
which is always less than zero.\ For example, $\mathcal{F}''(w_0)$ $=$ $-2N/3$ and $-N\epsilon$ respectively at small and large $k$ limit ($k/N$ $=$ $1$ $-$ $\epsilon$ with $\epsilon$ $\gtrsim$ $0$).\ 

Next for $\mathcal{F}''(w_{\pm 1})$ at the other two saddle points which we denote as $w_{\pm 1}$, we have from Eq. (\ref{roots}) with the "$-$" sign,
\bea
e^{iw_{\pm 1}}=\frac{-k/N-Q}{2(1+k/N)},~ e^{-iw_{\pm 1}}=\frac{k/N-Q}{2(1-k/N)},
\eea
where $w_{\pm 1}$ in general are complex.\ Setting $w_{\pm 1}$ $=$ $\pm\pi$ $+$ $iy_1$ with real $y_1$, we have $e^{-y_1}$ $\lesssim$ $1$ and $e^{-y_1}$ $\approx$ $1/2$ respectively when $k$ $\rightarrow$ $0$ and $k$ $\rightarrow$ $N$, suggesting $y_1$ $\gtrsim$ $0$ at small $k$ limit.\ Again we can use the above to reinterpret 
\bea
\cos w_{\pm 1}&=&\frac{(k/N)^2-Q}{2[1-(k/N)^2]},~ 1+2\cos w_{\pm 1}=\frac{1-Q}{1-(k/N)^2},\nonumber\\
\sin w_{\pm 1}&=&\frac{i}{2}\frac{k}{N}\frac{1-Q}{1-(k/N)^2}.
\eea
Now the second derivative of $\mathcal{F}(w_{\pm 1})$ becomes
\bea
\mathcal{F}''(w_{\pm 1})=-N\frac{Q[1-(k/N)^2]}{Q-1},\label{eq2}
\eea
which is again always less than zero, for example, $\mathcal{F}''(w_0)$ $=$ $-2N$ and $-2N/3$ respectively for small and large $k$ limit.\ 

Now we have located three saddle points, which are $w_0$ and $w_{\pm 1}$.\ The asymptotic form of the integral for $f^{(N)}_k$ in Eq. (\ref{integral}) can then be calculated using the contour in Fig. \ref{fig8}, which traverses through these three saddle points on the paths tangent to the real axis.\ Following the integration contour for Eq. (\ref{integral}), we have the asymptotic form $\bar f_k^{(N)}$ of $f^{(N)}_k$ as
\bea
\bar f_k^{(N)}=\frac{1}{2\pi}\left[\int_{D_{-1}}+\int_{C_{-1}}+\int_{C_0}+\int_{C_1}+\int_{D_1}\right]d\theta e^{\mathcal{F(\theta)}}.\nonumber\\
\eea
We obtain the contributions of $w_{0}$ in the path $C_{0}$ as
\bea
&&\frac{1}{2\pi}\int_{C_{0}} dwe^{\mathcal{F}(w_{0})} e^{\frac{1}{2}\mathcal{F}''(w_{0})(w-w_{0})^2},\nonumber\\
=&&\frac{e^{\mathcal{F}(w_{0})}}{2\pi}\int_{C_{0}} dw e^{-\frac{1}{2}|\mathcal{F}''(w_{0})|(w-w_{0})^2},\nonumber\\
=&&\frac{e^{\mathcal{F}(w_{0})}}{2\sqrt{\pi}\sqrt{-\mathcal{F}''(w_{0})/2}},\label{asymptotic}
\eea
where its next correction term is at least $\mathcal{O}(N^{-1/2})$ smaller.\ Since the integration path $C_{0(\pm 1)}$ follows the valley of steepest descent, the integration is dominated near the region of the saddle point $w_{0(\pm 1)}$, where we are also able to allow the boundary of $w$ to $\pm\infty$.\ Similarly for the paths $C_{\pm 1}$, we have the contributions from $w_{\pm 1}$,
\bea
&&\frac{e^{\mathcal{F}(w_{1})}}{2\pi}\left[\int_{C_{-1}} dw e^{\frac{1}{2}\mathcal{F}''(w_{-1})(w-w_{-1})^2}\right.\nonumber\\
&&\left. +\int_{C_{1}} dw e^{\frac{1}{2}\mathcal{F}''(w_1)(w-w_{1})^2}\right],
\eea
where $\mathcal{F}(w_{-1})$ $=$ $\mathcal{F}(w_{1})$.\ Let $w$ $=$ $w_{\pm 1}$ $+$ $x$ in the paths $C_{\pm 1}$ respectively, we have
\bea
&&\frac{e^{\mathcal{F}(w_{1})}}{2\pi}\left[\int_0^\infty dx e^{-\frac{1}{2}|\mathcal{F}''(w_{-1})|x^2}+\int_{-\infty}^0 dx e^{-\frac{1}{2}|\mathcal{F}''(w_1)|x^2}\right]\nonumber\\
&&=\frac{e^{\mathcal{F}(w_{1})}}{2\pi}\int_{-\infty}^\infty dx e^{-\frac{1}{2}|\mathcal{F}''(w_1)|x^2},\nonumber\\
&&=\frac{e^{\mathcal{F}(w_{1})}}{2\sqrt{\pi}\sqrt{-\mathcal{F}''(w_1)/2}},
\eea
where we have used $\mathcal{F}''(w_{-1})$ $=$ $\mathcal{F}''(w_{1})$.\ For the paths of $D_{\pm 1}$, though they do not follow the valleys of steepest descent, their contributions cancel with each other since $\mathcal{F}(w)$ $=$ $\mathcal{F}(w+2\pi)$.\

Finally we obtain the asymptotic form of $f^{(N)}_k$ in large $N$ limit as
\bea
\bar f^{(N)}_k=\frac{1}{2\sqrt \pi}\left[\frac{e^{\mathcal{F}(w_{0})}}{\sqrt{-\mathcal{F}''(w_0)/2}}+\frac{e^{\mathcal{F}(w_{1})}}{\sqrt{-\mathcal{F}''(w_1)/2}}\right].\nonumber\\
\eea
Before we write down the explicit form for the above, it is useful to derive
\bea
e^{\mathcal{F}(w_0)}=\left[\frac{1+Q}{1-(k/N)^2}\right]^N\left[\frac{Q-k/N}{2(1+k/N)}\right]^k,\label{eq3}
\eea
which become $3^N$ and $1/\epsilon^{N\epsilon}$ respectively for small and large $k$.\ Also we have 
\bea
e^{\mathcal{F}(w_{\pm 1})}=(-1)^{N+k}\left[\frac{Q-1}{1-(k/N)^2}\right]^N\left[\frac{Q+k/N}{2(1+k/N)}\right]^k,\nonumber\\\label{eq4}
\eea
which become $(-1)^{N+k}$ and $(3/4)^N2^{N\epsilon}/(-1)^{N\epsilon}$ respectively for small and large $k$.\ Inserting the functions of the saddle points from Eqs. (\ref{eq1}), (\ref{eq2}), (\ref{eq3}), and (\ref{eq4}), we have
\bea
\bar f^{(N)}_k=&&\left[\left(1+Q\right)^{N+\frac{1}{2}}\left(Q-\frac{k}{N}\right)^k\right.\nonumber\\
&&\left.+(-1)^{N+k}(Q-1)^{N+\frac{1}{2}}\left(Q+\frac{k}{N}\right)^k\right]\nonumber\\
&&\times\frac{\left[2\left(1+\frac{k}{N}\right)\right]^{-k}}{\sqrt{2NQ\pi}\left[1-\left(\frac{k}{N}\right)^2\right]^{N+\frac{1}{2}}}.\label{fNk}
\eea

\begin{figure}[b]
\centering
\includegraphics[width=8.5cm,height=4.5cm]{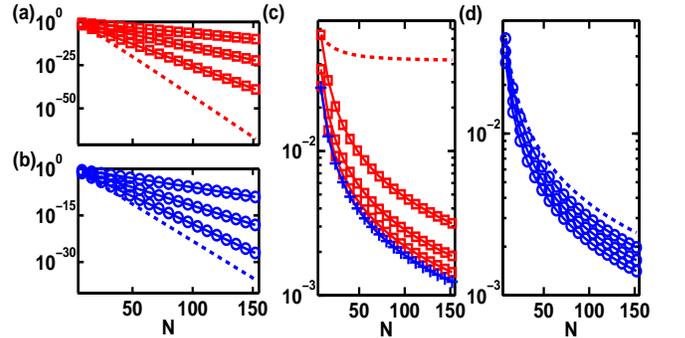}
\caption{(Color online) Exact spin function overlaps and their relative deviations from analytical derivations.\ Exact spin function overlaps, $(E,P_{12...j})_{+,0}$, for spin-plus and spin-$0$ components are plotted in (a) and (b) respectively, top to bottom, from $j/N$ $=$ $1/8$, $2/8$, and $3/8$ (solid), to $j$ $=$ $N/2$ $-$ $1$ (dash).\ The values decrease rapidly as $j$ increases.\ The relative deviations of the exact spin function overlaps $(E,P_{12...j})_{+,0}$ from $\bar f^{(N-j)}_{j,0}/\bar f^{(N)}_0$ are plotted for $j$ $=$ $1/8$, $2/8$, and $3/8$ (solid-$\square$ and $\circ$), and $j$ $=$ $N/2$ $-$ $1$ (dash), bottom to top, in (c) and (d) respectively.\ The deviations increase as $j$ increases.\ The relative deviations of $(E,E)_{+,0}$ from $\bar f^{(N-1)}_{1,0}/\bar f^{(N)}_0$ are denoted as $(+)$ in (c), which almost overlap with each other.\ $(E,E)_{+(0)}$ reaches $0.333(0.334)$ at $N$ $=$ $150$ and should approach $1/3$ for even larger $N$ as indicated by Eq. (\ref{largeN2}).}\label{fig9}
\end{figure}

We note that the main contribution in Eq. (\ref{fNk}) comes from the saddle point $w_0$.\ To have some estimates of Eq. (\ref{fNk}), we have
\bea
\bar f^{(N)}_{k\rightarrow 0}=\frac{3^N}{2\sqrt{\pi}\sqrt{N/3}},~ \bar f^{(N)}_{k\rightarrow N}=\frac{1}{\sqrt{2\pi N\epsilon}\cdot \epsilon^{N\epsilon}}.
\eea
From the above result at small $k$ and according to Eq. (\ref{relation}), we can show that
\bea
&&\frac{(E,P_{12...j+1})_{+(0)}}{(E,P_{12...j})_{+(0)}}\bigg|_{N\rightarrow\infty} = \frac{1}{3},\nonumber\\
&&(E,E)_+\big|_{N\rightarrow\infty} =(E,E)_0\big|_{N\rightarrow\infty}  = \frac{1}{3},\label{largeN2}
\eea
which respectively indicates one third decrease for the spin function overlaps when $j$ increases by one, and the populations of $N_+$ and $N_0$ coincide in large $N$ limit.\ In the large $N$ limit, despite the constraint $S_z$ $=$ $0$, the probability of finding a particle in any of the spin states, "$+$", "$0$", or "$-$", is $1/3$ and is irrespective of the spins of the other particles (if $j$ $\ll$ $N$).\ This decrease in spin function overlaps also reflects on the exponential decay in spatial correlations discussed in Sec. III. A.

To have some estimates of the spin function overlaps and their asymptotic forms in large $N$ limit, in Fig. \ref{fig9} we compare $(E,P_{12...j})_{+,0}$ of Eqs. (\ref{EP_p}) and (\ref{EP_0}) with $\bar f^{(N-j)}_{j,0}/\bar f^{(N)}_0$ from Eq. (\ref{fNk}).\ In Figs. \ref{fig9}(a) and (b), the values of spin function overlaps decrease rather fast in logarithmic scales as $j$ increases, while $(E,P_{12...j})_{0}$ is always larger than $(E,P_{12...j})_+$.\ As a comparison, we define the relative deviations as
\bea
\left|\frac{(E,P_{12...j})_{+,0}-\bar f^{(N-j)}_{j,0}/\bar f^{(N)}_0}{(E,P_{12...j})_{+,0}}\right|,
\eea 
which we show in Figs. \ref{fig9}(c) and (d), indicating of a good asymptotic form from our derivations for small $j$ in large $N$ limit.\ However a slow decay in the relative deviation of $(E,P_{12...N/2-1})_+$ to $\bar f^{(N/2+1)}_{N/2-1}/\bar f^{(N)}_0$ in (c) shows the worst case in the asymptotic form.\ It is due to a rather small $\mathcal{F}''(w_0)$ $\rightarrow$ $-N\epsilon$ in Eq. (\ref{eq1}) when we set $k/N$ $=$ $1$ $-$ $\epsilon$ with a small value of $\epsilon$, which makes the method of steepest descent less accurate unless we go to $N$ $\rightarrow$ $\infty$.


\end{document}